\documentclass[a4paper,fleqn,usenatbib]{mnras}
\usepackage{newtxtext,newtxmath}
\usepackage[T1]{fontenc}
\usepackage{ae,aecompl}
\usepackage{graphicx}	
\usepackage{amsmath}	
\usepackage{amssymb}	

\title[Hoag's jellyfish galaxy]{GASP V: Ram-pressure
  stripping of a ring Hoag's-like galaxy in a massive cluster}

\author[A. Moretti et al.]{
A. Moretti,$^{1}$\thanks{E-mail: alessia.moretti@oapd.inaf.it}
B.M. Poggianti,$^{1}$ M. Gullieuszik,$^{1}$ M. Mapelli,$^{1}$ Y.L.
Jaff\'e,$^{2}$ J. Fritz,$^{3}$
\newauthor A. Biviano,$^{4}$ G. Fasano,$^{1}$
D. Bettoni,$^{1}$ B. Vulcani,$^{1,5}$ and M. D'Onofrio$^{6}$
\\
$^{1}$ INAF-Astronomical Observatory of Padova   
vicolo dell'Osservatorio 5   
35122 Padova, Italy \\
$^{2}$ European Southern Observatory, Alonso de Cordova 3107,
  Vitacura, Casilla 19001, Santiago de Chile, Chile \\
$^{3}$ Instituto de Radioastronomia y Astrofisica, UNAM, Campus 
  Morelia, A.P. 3-72, C.P. 58089, Mexico \\
$^{4}$ INAF-Osservatorio Astronomico di Trieste, via G.B. Tiepolo 11,
  34131 Trieste, Italy \\
$^{5}$ School of Physics, The University of Melbourne, Swanston St \&
  Tin Alley Parkville, VIC 3010, Australia \\
$^{6}$ Department of Physics and Astronomy, University of Padova,
  vicolo dell'Osservatorio 5, 35122 Padova, Italy
}

\date{Accepted XXX. Received YYY; in original form ZZZ}

\begin{document}
\label{firstpage}
\pagerange{\pageref{firstpage}--\pageref{lastpage}}
\maketitle

\begin{abstract}
Through an ongoing MUSE program dedicated to study gas removal
processes in galaxies (GAs Stripping Phenomena in galaxies with MUSE, GASP), we
have obtained deep and wide integral field spectroscopy of the galaxy
JO171. 
This galaxy resembles the Hoag's galaxy, one
of the most spectacular examples of ring galaxies, characterized by a
completely detached ring of young stars surrounding a central old
spheroid.  At odds with the isolated Hoag's galaxy, JO171 is part of a
dense environment, the cluster Abell 3667, which is causing gas
stripping along tentacles.
Moreover, its ring counter-rotates with respect to the central
spheroid.  The joint analysis of the stellar populations and the
gas/stellar kinematics shows that the origin of the ring was not due to
an internal mechanism, but was related to a gas accretion event 
that happened in the distant past, prior to accretion onto Abell 3667,
most probably within a filament. 
More recently, since infall in the 
cluster,
the gas in the ring has been stripped by ram-pressure,
causing the quenching of star formation in the stripped half of the ring.
This is the first observed case of ram pressure stripping in action in
a ring galaxy, and MUSE observations are able to reveal both of the events (accretion
and stripping) that caused dramatic transformations in this galaxy. 
\end{abstract}




\section{Introduction}

The current understanding of galaxy formation and evolution links
galaxy properties to both  the original conditions in 
which galaxies formed, and the environment in which they are embedded.

According to the current cosmological paradigm, most galaxies in the
Universe undergo at least an episode of accretion, 
or merger, or collision. Observational signatures of such events are
transient features that make galaxies morphologically 
peculiar for a relatively short time, typically for a few tens of Myr to
a few Gyr. A peculiar morphology is therefore a transient 
phase in the life of a galaxy.

Peculiar galaxies, such as ring galaxies and galaxies with
counter-rotating components, can 
offer the key to 
understand the mechanisms at play in galaxy formation, revealing for
example whether gas accretion 
and/or mergers are playing a major role in shaping galaxies properties.

Both rings and counter-rotation can be explained in terms of internal or external mechanisms.
In fact, the presence of stellar and gaseous rings around galaxies can
be due to different effects \citep{Schweizer+1987}. Internal mechanisms, such as slow
secular evolution, can produce
rings as a consequence 
of the formation of a bar that dynamically induces the accumulation of material in
correspondence to the Lindblad resonances, i.e. the regions where
radial gas inflows are slowed down \citep{Buta1993,Heller1996}.
External mechanisms are related to environmental processes, 
such as major or minor mergers, or gas accretion
either of low metallicity circumgalactic gas \citep{Spavone2013} or from a satellite galaxy \citep{ButaCombes1996}.

Moreover, head-on collisions between a disk galaxy and a companion galaxy can
produce the so-called collisional ring galaxies (the most famous
example being the Cartwheel galaxy).  In this case, the passage of the
companion through the disk induces a perturbation in the orbits of the
disk stars which propagates as a density wave \citep{Mapelli+2008}. This implies that
collisional rings are composed of stars that were already present in
the disk galaxy, with no significant accretion from the companion
galaxy.  Moreover, the ring preserves the rotation of the original
disk.

The idea of external gas accretion, while initially ruled out to
explain ring galaxies due to the low frequency of isolated gas clouds
with respect to the number of observed galaxies \citep{Appleton1996}, has been recently
reconsidered in the cold mode of gas accretion \citep{Maccio2006}.
Evidence for this mechanism is
corroborated by observations of counter-rotation in some ringed
galaxies (e.g. IC 2006 and NGC 77427, \citealt{Silchenko2006}).  In
particular, this scenario
has been invoked to explain the origin of polar ring galaxies,
i.e.  galaxies that have two kinematically distinct components,
inclined by almost 90 degrees to each other \citep{Spavone2010}, as
well as
the origin of rings highly inclined with respect to
the host galaxy disk in general \citep{Spavone2013}. 

Some of the mechanisms proposed for the formation of ring galaxies
have been invoked also for the existence of counter-rotating
components in galaxies \citep{Bettoni+2001,Mapelli+2008}.
Theoretical studies have shown that both a prolonged gas infall and a
merger with a dwarf galaxy can lead to the observed counter-rotation
\citep{Jore1996,Thakar+1996}.  In particular, in the merger
case, the end product is an undisturbed optical morphology if the merger
occurred more than 1 Gyr before the observations and, therefore, it
would be accompanied by >1 Gyr old stars in the counter-rotating
component.  

The formation of counter-rotating gaseous disks seems to
be favored in gas-poor systems, like S0 galaxies, while in spiral
disks, hosting large amounts of gas co-rotating with the stellar
component, is less frequent \citep{Bettoni2014}. In fact, a
counter-rotating gaseous disk will be observed only if the mass of the
newly supplied gas exceeds that of the pre-existing one \citep{Lovelace+1996}.  As for the stellar component, the counter-rotating
components detected in samples of S0 galaxies \citep{Johnston+2012,Coccato+2015,Katkov+2013} have younger stellar
populations compared to the main stellar disks, and their ionized gas
rotates in the same direction as the secondary stellar components,
i.e. it also counter-rotates with respect to the main disk. The
youngest counter-rotating stars, therefore, should have been formed
from the externally accreted gas. Evidence in favor of this
interpretation comes also from recent results obtained with integral
field spectroscopy of blue galaxies in
the MaNGA survey \citep{Chen+2016}.

One clear example of a non-collisional ring galaxy is the
Hoag's object, an isolated galaxy at redshift 0.04 consisting of a central
spheroid surrounded by a completely detached ring of young stars and
gas.  The absence of a central bar, as well as of an inclined disk,
together with the strong contribution of the external ring to the
total light of the galaxy (at least 20\% in the case of the Hoag's
object) are its main characteristics, and led \citet{Schweizer+1987}
to use these features to define a class of objects, called Hoag-type
galaxies, which are neither obviously barred nor obviously inclined
disks, and which have outer rings containing a significant fraction of
the total luminosity.  The presence of a dissolved bar was
invoked as the origin of the ring in the Hoag's galaxy by \citet{Brosch1985}. However, subsequent
studies revealed that the core of this peculiar galaxy
is a spheroidal object, and not a disk, thus favoring a scenario where
the ring was formed due to an accretion event 2-3 Gyr ago \citep{Schweizer+1987,Finkelman+2011}.  Integral field spectroscopy of
the Hoag's galaxy \citep{Finkelman+2011}  confirmed that the
central spheroid is consistent with being a classical bulge, typical
of an elliptical galaxy.  Its age could be as old as 10 Gyr, while the
ring's age is younger than 2 Gyr.  Stars in the central spheroid and
in the ring share the same kinematics, i.e. rotate in the same direction.

In what follows we present MUSE integral-field spectroscopy of JO171, 
a peculiar galaxy that resembles the Hoag's galaxy 
and is undergoing strong ram pressure stripping in a massive galaxy
cluster. We will
discuss how its stellar and gas properties can be explained,
 searching for the origin of its doubly peculiar
characteristics. After having presented our
observations and methods (\S2), we will show the results regarding
the structure of JO171 (\S3.1), its gas and stellar kinematics
(\S3.2), stellar population ages (\S3.3) and environment (\S3.4),
discussing the formation and evolution of this galaxy in \S4 and summarizing
our findings in \S5.
Throughout this paper, we use a standard concordance cosmology with
$H_0 = 70 \, \rm km \, s^{-1} \, Mpc^{-1}$, ${\Omega}_M=0.3$
and ${\Omega}_{\Lambda}=0.7$ and a \citealt{Chabrier2003} IMF.

\section{Observations and methods}

In optical images, JO171 (RA=20:10:14.6, DEC=-56:38:20.7) shows a 
disturbed morphology with one-sided tails 
suggestive of stripping by the Intra Cluster Medium (ICM). It was therefore been 
tagged as a candidate ``jellyfish'' galaxy  by \citet{Poggianti+2016}. Jellyfish galaxies are those 
exhibiting ongoing star formation occurring in tails of stripped gas
\citep{Smith+2010}. 

JO171 is a member  of the cluster Abell 3667 (A3667), surveyed by the WIde 
Field Nearby Galaxy-Cluster Survey (WINGS, \citealt{Fasano+2006,Moretti+2014}) and its extension OmegaWINGS \citep{Gullieuszik+2015,Moretti+2017}. Its WINGS ID is WINGSJ201014.69-563830.1 and its 
spectroscopic redshift is 0.052529. At this redshift 1 arcsec corresponds to 1.01 kpc.

This galaxy is part of the GASP survey (Gas Stripping Phenomena with
MUSE, \citealt{gaspI}), dedicated to study jellyfish galaxies across a wide range of
galaxy masses and environments. GASP uses the Integral Field
Spectrograph MUSE at the Very Large Telescope (VLT) \citep{Bacon+2010}, which is capable
of obtaining a spectrum for each of the $\sim$90000 spaxels covering the 1\arcmin $\times$ 1\arcmin field of view.
The MUSE datacube has a spectral sampling of 1.25 \AA/pix, a spatial
sampling of 0.2 arcsec/pix and a spectral range between 4500 and 9300
\AA\, with a spectral resolution of $\sim 2.6 \AA$.
JO171 was observed with MUSE on 13 May 2016, with a seeing of 0.99 arcsec, for a
total exposure time of 2700sec split into three 675 sec exposures.
The data were reduced using the ESO MUSE pipeline following the
standard GASP procedure fully described in \citet{gaspI}.

From the MUSE datacube we measured the gas kinematics by fitting the
emission lines ($\rm H\alpha$, $\rm H\beta$, [NII], [OIII]), using the KUBEVIZ
software \citep{Fossati2016}. The velocity of each spaxel in the frame is derived with
respect to a given redshift, that is the redshift of the
spheroid. Among all the emission lines used by KUBEVIZ the most
prominent ones are the triplet around $\rm H\alpha$ comprising also the Nitrogen
lines. This triplet has been fitted with a common continuum and
variable fluxes. The continuum is calculated as the mean value
calculated inside two windows redwards and bluewards of each line
between 80 and 200 \AA, and using values between 40th and 60th
percentiles.  The ratio between the two [NII] and [OIII] lines has
been kept constant in the fit assuming the ratios given in \citet{Storey2000}.  Measurements have been made on a mean 
filtered 5x5 pixels kernel datacube, in order to maximize the signal to noise.	

We determined the stellar kinematics using the Penalized Pixel-Fitting
(pPXF) software \citep{CE2004} on spatially binned spectra. For the binning we used
a Voronoi tessellation algorithm \citep{CC2003} that analyzes the white light image
extracted from the original datacube and bins spectra on the basis of
their signal-to-noise (S/N). We imposed a S/N=10.  The derived spectra have
been fitted with the \citet{MILES} stellar population templates. In particular we
used Single Stellar Populations (SSP) with metallicities that range
from [M/H]=-1.71 to [M/H]=0.22 and ages from 1 to 17.78 Gyr,
calculated with the \citet{Girardi+2000} isochrones.  

We used our spectrophotometric code SINOPSIS, recently modified to be able to work with
observed datacubes \citep{gaspIII}.  The code fits the observed spectra with a combination of SSPs from Charlot \&
Bruzual (in prep.) calculated with a \citet{Chabrier2003} Initial Mass
Function (IMF) between 0.1 and 100 solar masses. The range in
metallicities goes from Z=0.0001 to Z=0.04. 
On this set of SSPs
nebular emission has been added.  The code is also able to properly
handle spectra with superimposed gas and stellar components with
different velocities, as the two redshifts are given as inputs to the
code.
The SINOPSIS code gives as output a best fit model cube, as well as a
stellar only model cube, that is then subtracted from the original one
to obtain an emission only cube, useful for the determination of
emission line fluxes corrected for stellar absorption.  Finally, it
gives stellar masses, star formation histories and stellar ages (both
luminosity and mass weighted) for each spaxel.  Due to the
characteristics of the SINOPSIS code, we calculated the total mass in
a given spaxel as the sum of the masses in four main age bins, but
neglecting the mass of stars older than 0.57 Gyr when the spectra had
a global S/N lower than 3. In these cases, in fact, the code is forced
to find a not meaningful solution that includes a low percentage of
old stars. The contribution of young stars, instead, is taken
into account, given the fact that it is estimated from the emission
lines, which are more trustable features.

The gas metallicity has been derived using the line flux
measurements made on the emission-only and dust-corrected spectra of
the spaxels where the ionization is typical of HII regions. In
particular the internal extinction correction due to the galaxy dust
has been taken into account measuring the Balmer decrement in each
spaxel, and assuming an intrinsic $\rm H\alpha$/$\rm H\beta$=2.86 and the \citet{CCM1989} extinction law .  We then used the pyqz code \citep{Dopita+2013} to derive both the
ionization parameter q and the gas metallicity through the comparison
with a set of photo-ionization models (MAPPINGS IV). In particular, we
used the [NII]6583/[SII]6716,6731 vs [OIII]5007/[SII]6716,6731
indicator, given that the [OII]3727 doublet lies beyond the MUSE spectral
range, and that $\rm H\beta$ is prone to uncertainties in the subtraction of the
stellar contribution.  The systematic errors introduced by modeling 
inaccuracies are usually estimated to be $\sim$0.1-0.15 dex, whereas 
discrepancies of up to 0.2 dex exist among the various calibrations
based on photoionization models \citep{Kewley2008}.

\section{Results}

\subsection{JO171 structure}
\begin{figure*}
\centerline{\includegraphics[width=3.5in]{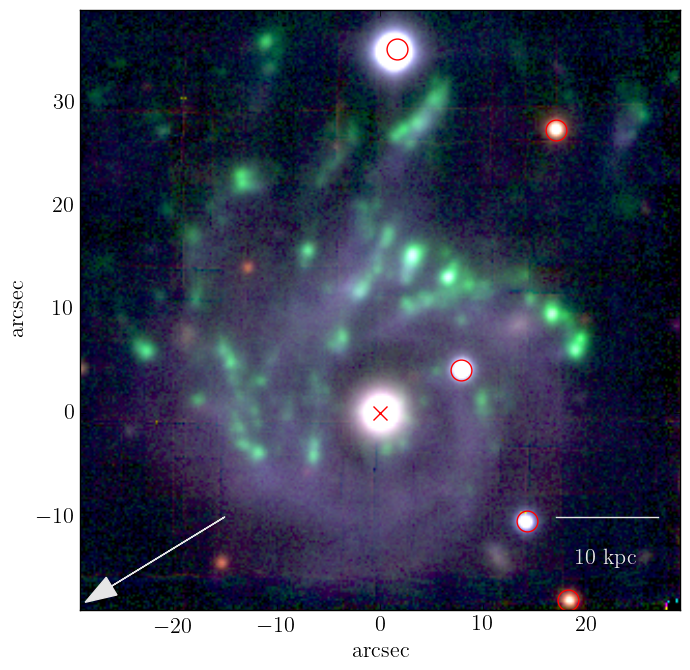}\includegraphics[width=3.5in]{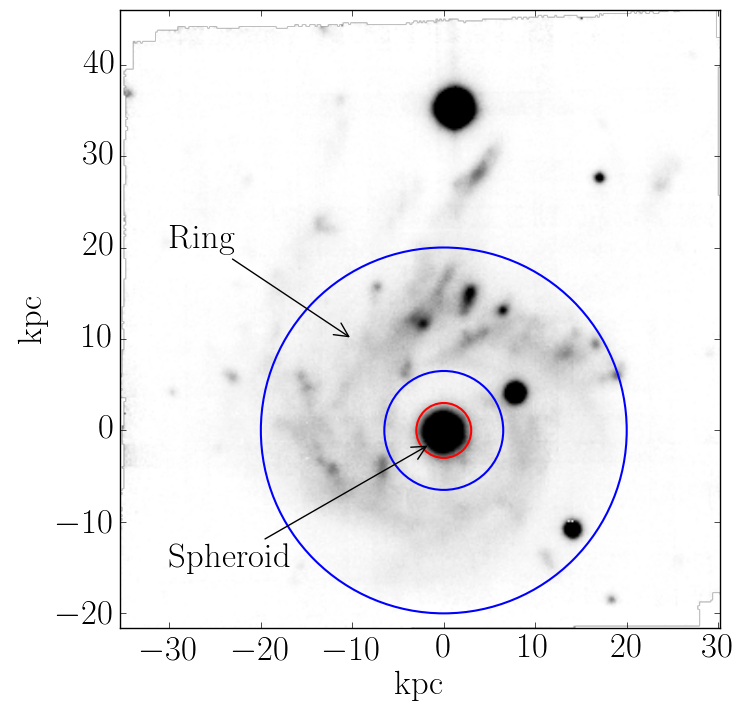}}
    \caption{Left. RGB image of JO171 extracted from the MUSE datacube. The 
      three bands correspond to the integrated flux in the Johnson I 
band (in grey colour), the integrated flux inside a 40 \AA-wide band 
centered on $\rm H\alpha$ (thus including both the $\rm H\alpha$ emission and the continuum 
under $\rm H\alpha$, here in green color) and the integrated flux in the Johnson 
B band (blue in the image). Red circles indicate foreground stars that 
have been masked for the purposes of our analysis. The superimposed 
scale shows that 1\arcsec corresponds to 1.01 kpc at the galaxy 
redshift. North is up, East to the left. The arrow indicates the 
direction towards the Brightest Cluster Galaxy. Right. A grayscale 
image of the galaxy with superimposed a red circle enclosing the 
spheroid ($R_e$), and blue circles enclosing the ring (see text).}
    \label{fig:rgb}
\end{figure*}

From the reduced datacube, we extracted images in three bands, namely
the I band, the $\rm H\alpha$ emission including the underlying continuum in a 40
\AA-wide band and the B band, and constructed the RGB image shown in
the left panel of Fig.\ref{fig:rgb}.  A grayscale image is shown in the right
panel. There are clearly two prominent structures: a central round
spheroid and an external ring detached from the spheroid. The external
ring shows a spiral-like pattern, more evident than in other Hoag-type
galaxies. The ring is traced by the stellar continuum (in
grey/white in Fig.\ref{fig:rgb}) and only his northern half has prominent
emission lines (in green in Fig.\ref{fig:rgb}). The central spheroid lacks
ionized gas, but for a few small blobs seen south and west of the
spheroid.  Figure\ref{fig:rgb} also shows clear signatures of extended tails of
ionized gas towards the north, in the direction opposite to the
cluster center, confirming that JO171 is a
jellyfish galaxy, whose disturbed morphology is linked
with the environmental conditions.

\begin{figure*}
\centerline{\includegraphics[width=3.5in]{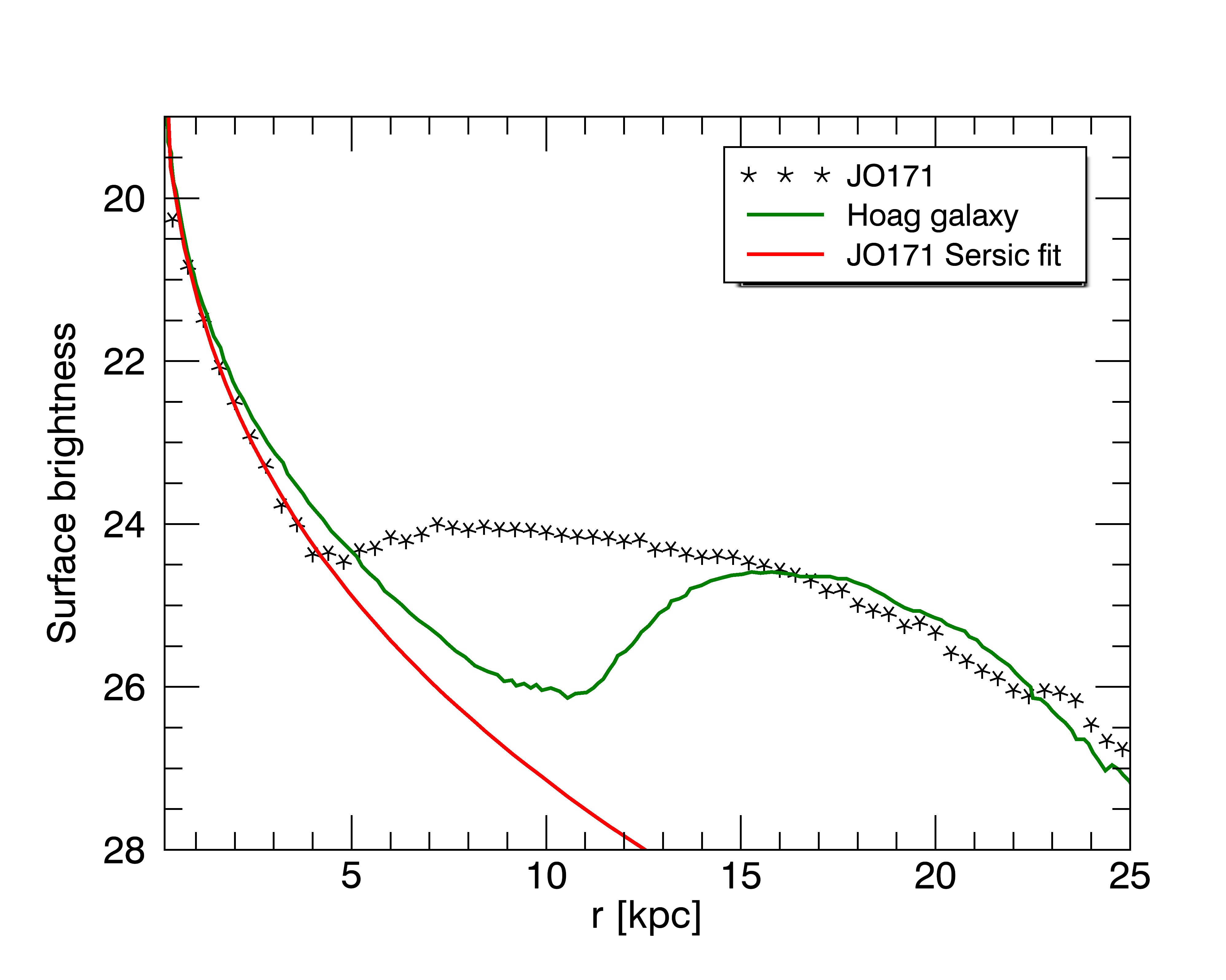}\includegraphics[width=3.5in]{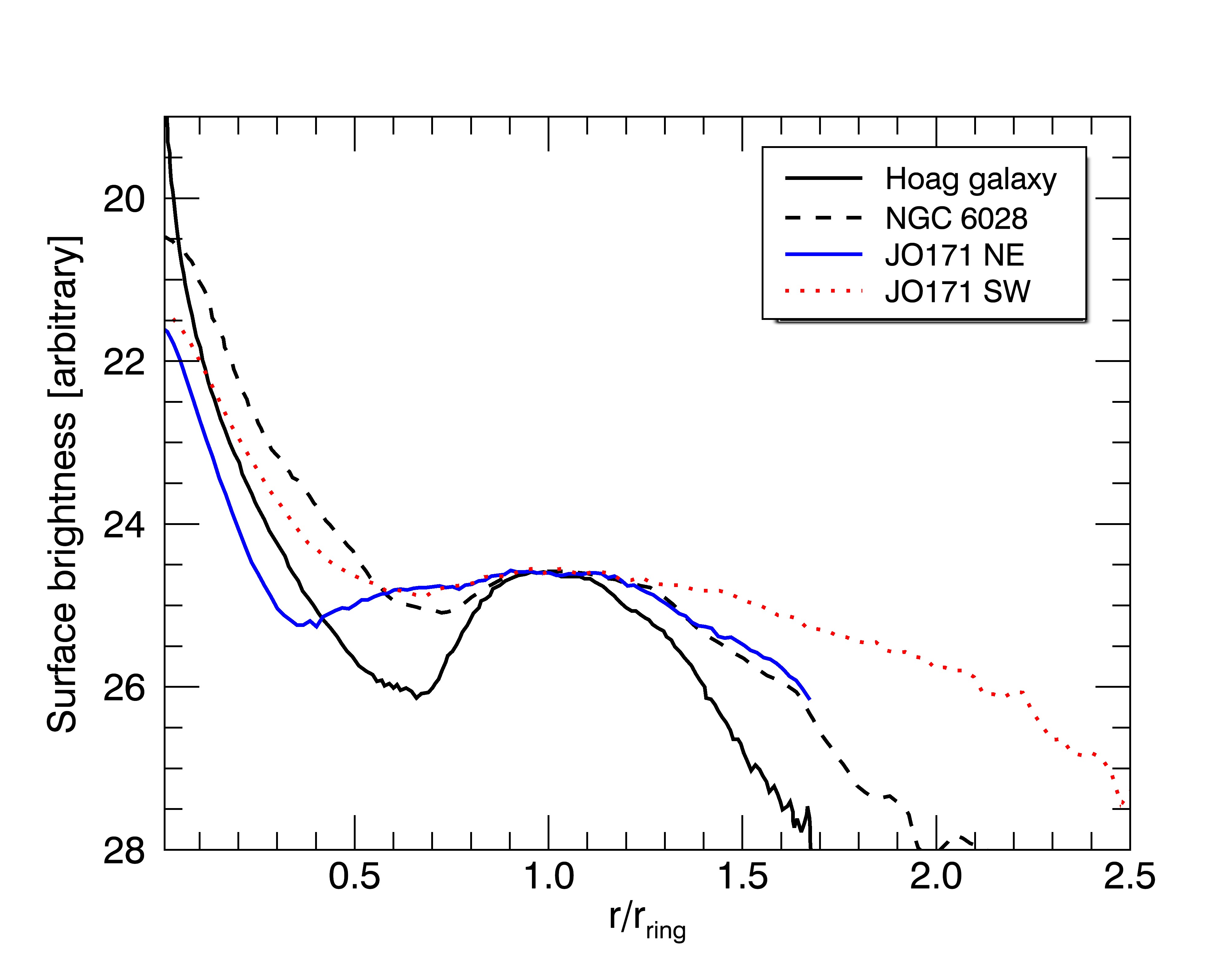}}
    \caption{Left. Surface brightness profile of JO171 in mag$~\rm arcsec^{-2}$,
      with superimposed the Sersic fit of the central spheroid in red. 
    The green line is the Hoag's galaxy profile. Right. The blue and 
    red profiles are the 180 degrees averaged profiles of JO171 
    north-east and south-west hemispheres, respectively. In black, 
   the surface brightness profiles of the Hoag's galaxy (continuous line) and of the Hoag-type galaxy NGC 6028 (dashed line).}
    \label{fig:profile}
\end{figure*}

The left panel of Fig.\ref{fig:profile} shows the JO171's B band surface brightness profile 
together with the
Hoag's galaxy surface brightness profile (in green, \citealt{Finkelman+2011}). The JO171 profile has been obtained averaging over all 
directions after having masked both the stars and the
regions characterized by very peaked star formation. For the sake of 
comparison, the surface brightness profiles have been normalized to the same central surface brightness.
We also derived the surface brightness profiles in
other bands (SDSS u-band from OmegaCAM images (D'Onofrio et al. in
prep.),  continuum under $\rm H\alpha$ and Cousin's I-band from the
MUSE data, Johnson K-band from VIRCAM archive data) and found that the JO171 ring becomes
fainter going towards longer wavelengths, showing that its light
comes primarily from young stars.

The structural parameters have been measured using a bi-dimensional fit using the GALFIT tool \citep{Peng+2010}.  The central
spheroid has been fitted with a Sersic law that has two
free parameters: the Sersic index n and the effective radius $R_e$ that
turned out to be 2.96$\pm$0.10 and 1.39$\pm$0.06 kpc, respectively.  
The Sersic fit of the central spheroid is shown in red in the left
panel of Fig.\ref{fig:profile}.
The surface brightness analysis led us to better define the two structures composing JO171: in
the following, we call {\it spheroid} the
region within a radius of 3 kpc, and {\it ring} the region extending from 6.5
kpc (i.e. the distance at which the surface brightness profile starts
to rise) out to the average distance where counts reach $3\sigma$ 
above the background, that is at $\sim$20 kpc.
We have marked the spheroid's effective radius and the limits of the two
regions in the right panel of Fig.\ref{fig:rgb}. 

We integrated the MUSE spectra inside these two regions and fit them
with the SINOPSIS code described in \S2.
The total galaxy mass of JO171 is $3.39\pm0.5 \times
10^{10}M_{\odot}$, with 56\% of stars in the central spheroid, and 32\% in the
ring. The remaining mass is locked in the region between the spheroid 
and the ring (4\%) and in the stripped tails ($\sim$8\%). 

\begin{figure}
\centerline{\includegraphics[width=0.45\textwidth]{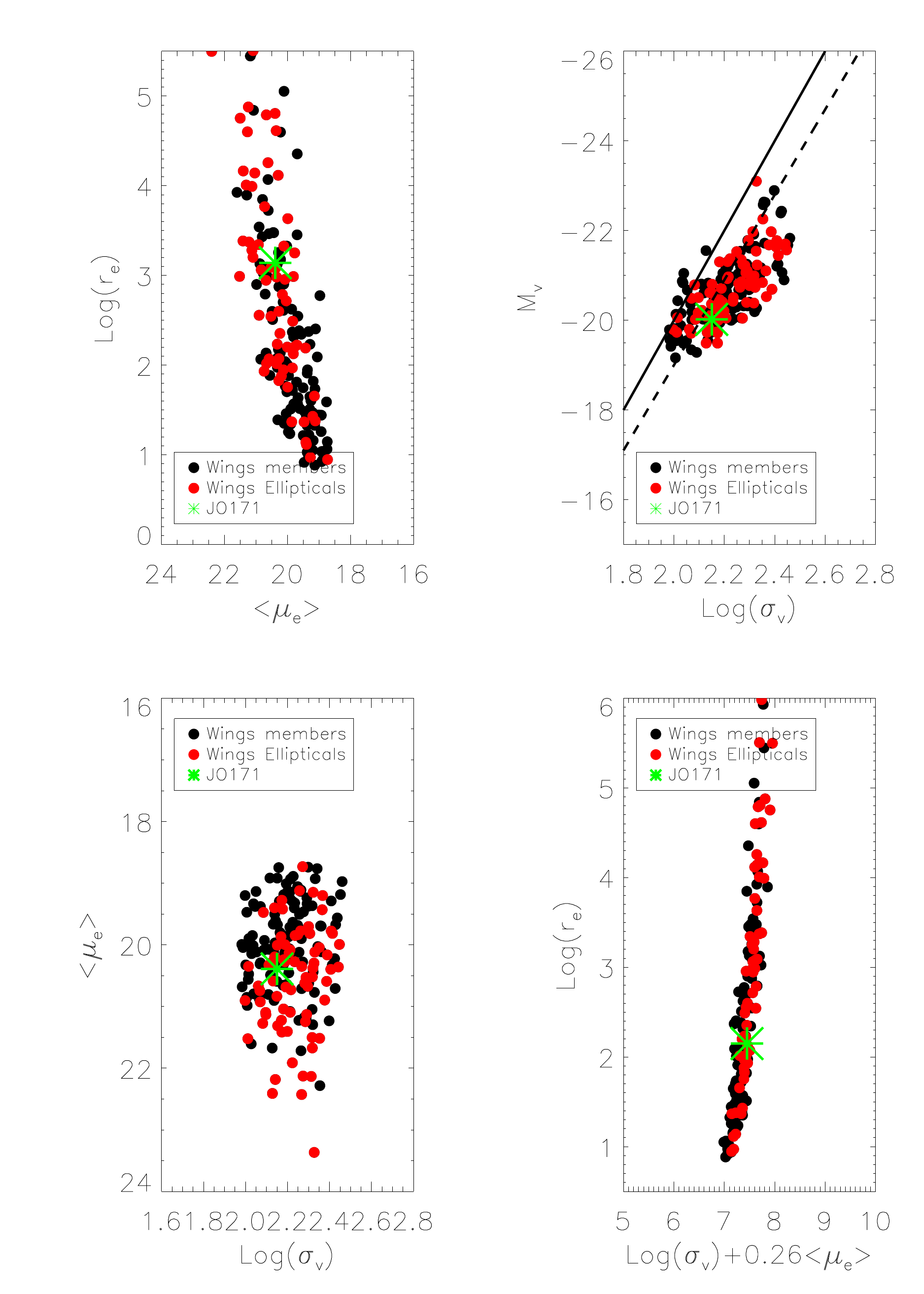}}
    \caption{Fundamental Plane projections for WINGS cluster 
      galaxies. In all panels the dots are spectroscopically confirmed
      cluster members, ellipticals (red) and S0s (black). The green asterisk is JO171. In the 
      Faber -Jackson plane 
 the continuous line shows the $L \propto \sigma^4$ relation, the dashed line a $L \propto \sigma^{3.8}$ relation.}
    \label{fig:fp}
\end{figure}

Comparing with all WINGS clusters members with a known velocity
dispersion (Fig.\ref{fig:fp}), we find that the position of the JO171 spheroid
in the various projections of the Fundamental Plane is perfectly in
agreement with the locus 
occupied by elliptical/early-type
galaxies, leading us to conclude that this is not a
pseudo-bulge. Moreover, the spheroid characteristics ($v_{max}=60 \rm km \,
s^{-1}$, $\sigma_c \sim 120 \rm km s^{-1}$, $e=0.05$, where $\sigma_c$ is the central velocity dispersion and the ellipticity is the average ellipticity within the spheroid)
demonstrate that, similarly to the Hoag's central spheroid, this is a
fast rotator, which indicates
that the merging hypothesis for its formation is very unlikely.

The structure of JO171 looks similar to the Hoag's object: the
two galaxies have a central spheroid of similar luminosity (-20.8,
-20.7 in the B band, respectively), and the same effective radius of
$\sim$2 kpc \citep{Finkelman+2011}. Both spheroids are almost round (ellipticity e=0.03 for
Hoag's, 0.05 for JO171).
The ring extends
approximately between 12 and 25 kpc in the Hoag's galaxy, and from 6.5 to 20 kpc in
JO171. We note that if we considered a ring extending out to 25 kpc
the results that we present in the following would not be altered.
The mass of stars in the ring of Hoag's galaxy was estimated from
its luminosity and simple assumptions on the stellar populations by
\citet{Finkelman+2011}, and is $\sim 3 \times 10^9 M_{\odot}$,
smaller than that of JO171 ($1.1 \times 10^{10} M_{\odot}$). 

At odds with the Hoag's galaxy, JO171 shows extended tails towards the
north 
characterized by ionized
gas and ongoing star formation: while the Hoag's galaxy is isolated, JO171 belongs to a 
very active and rich galaxy cluster whose intracluster medium exerts 
a ram pressure force that creates the tails of stripped gas to the north.

As a consequence, its ring is clearly asymmetric in terms of the stellar populations/gas
properties: the stripping of gas by ram pressure has
caused the south-western part of the galaxy to be devoid of gas and,
as we will see in \S3.3, of ongoing star formation.  This effect should be
considered when comparing its surface brightness profile with the
Hoag's galaxy and other Hoag-type galaxies. Ideally, we would
like to analyze the JO171's profile {\it before} gas stripping began,
and one way to approximate this situation is to consider the half of JO171 that has not
been quenched/gas stripped yet, i.e. the north-eastern part.

In order to evaluate the differences between the
star-forming/gas-rich half of JO171's ring and the passive/gas-poor half of
it, we draw a line inclined by 65 degrees N-W and calculate the 180
degrees average surface brightness profiles, that are shown in
the right panel of Fig.\ref{fig:profile}, together with the Hoag's galaxy and NGC
6028 profiles (black continuous and dashed lines, respectively): the blue
line shows the star-forming half of the ring, the red line the passive
portion of it.  In this figure we rescaled the radius to the ring radius and the 
surface brightness to the Hoag's galaxy ring surface brightness, in
order to evaluate the magnitude dip between the spheroid and the ring in the
different galaxies more easily. 

It is clear from this figure that the profiles of the star-forming and
passive halfs of JO171 are
different, and in particular that the north-east part of the ring
shows a deeper gap between the spheroid and the ring.  
The magnitude difference is maximum for the Hoag's galaxy (1.5
magnitudes), and is minimum for the passive part of JO171's ring (0.3
magnitudes). NGC 6028 shows a dip of 0.5 magnitudes, while the star
forming part of JO171's ring reaches 0.7 magnitude difference.

To conclude, JO171, as the Hoag's object and Hoag-type galaxies, is
composed of two structures, a central round spheroid and a
detached surrounding ring, whose origin will be investigated in the
following sections.

\subsection{Gas and stellar kinematics}

From the MUSE datacube we measured the gas kinematics from the 
$\rm H_{\alpha}$ emission line and the stellar kinematics
as described in \S2.
Formal errors in the
velocity estimation of the gas component are below 3 $\rm km \, s^{-1}$ for most of
the spaxels.
The ring and the central spheroid share the same systemic velocity
confirming that the two components belong to a unique system.

\begin{figure*}
\centerline{\includegraphics[width=3.5in]{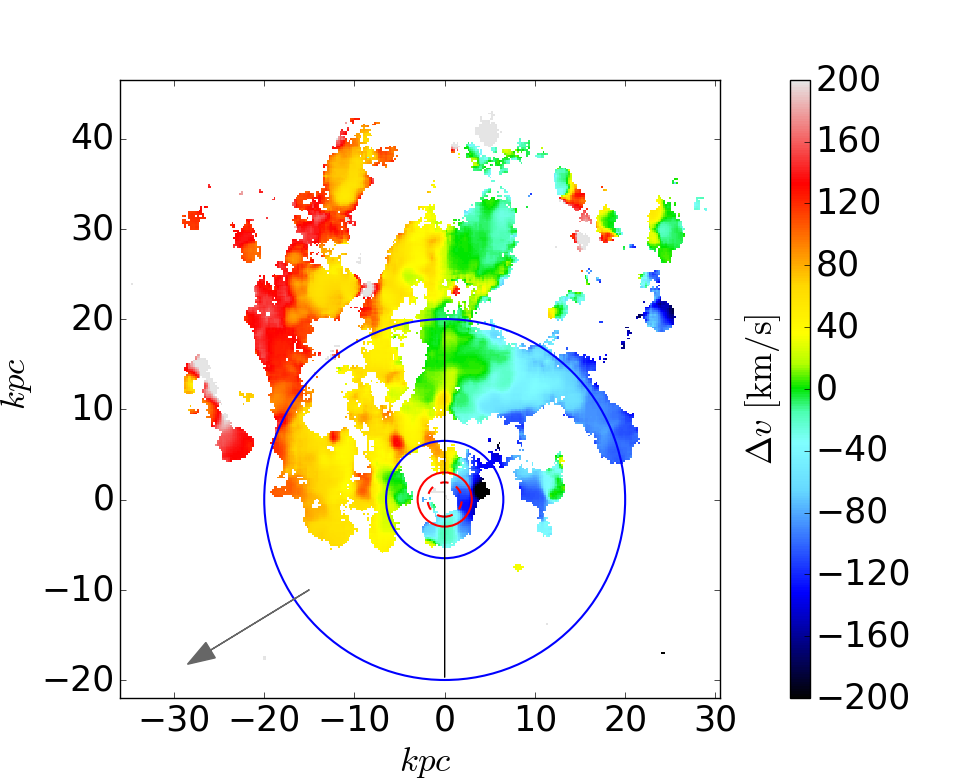}\includegraphics[width=3in]{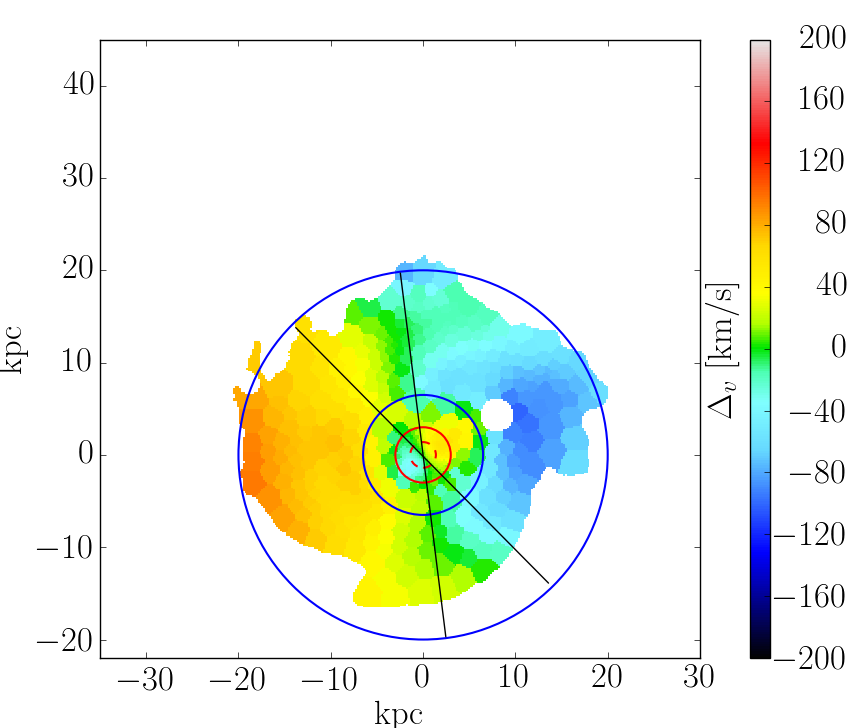}}
\centerline{\includegraphics[width=3in]{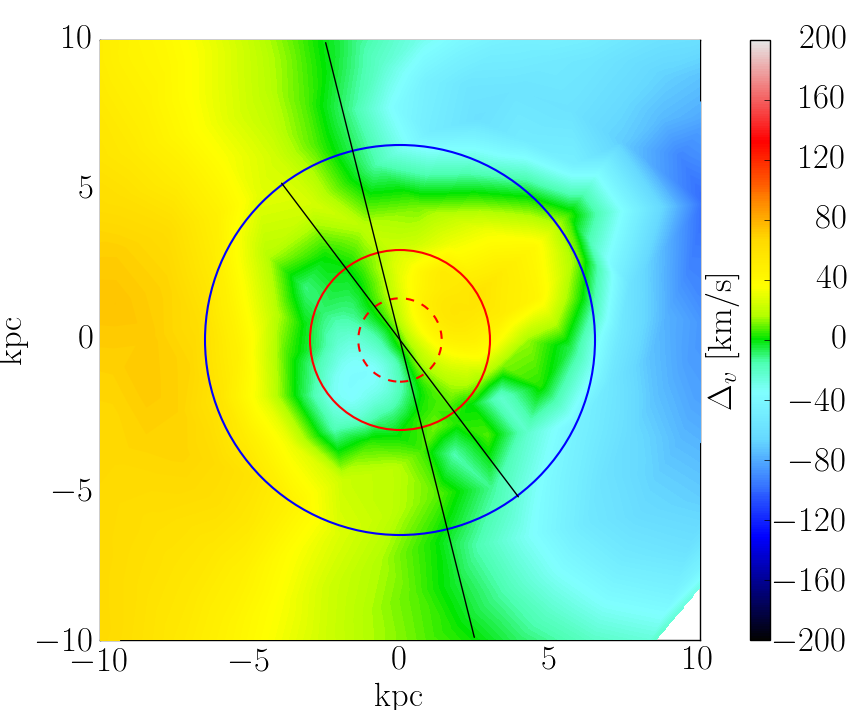}\includegraphics[width=3.5in]{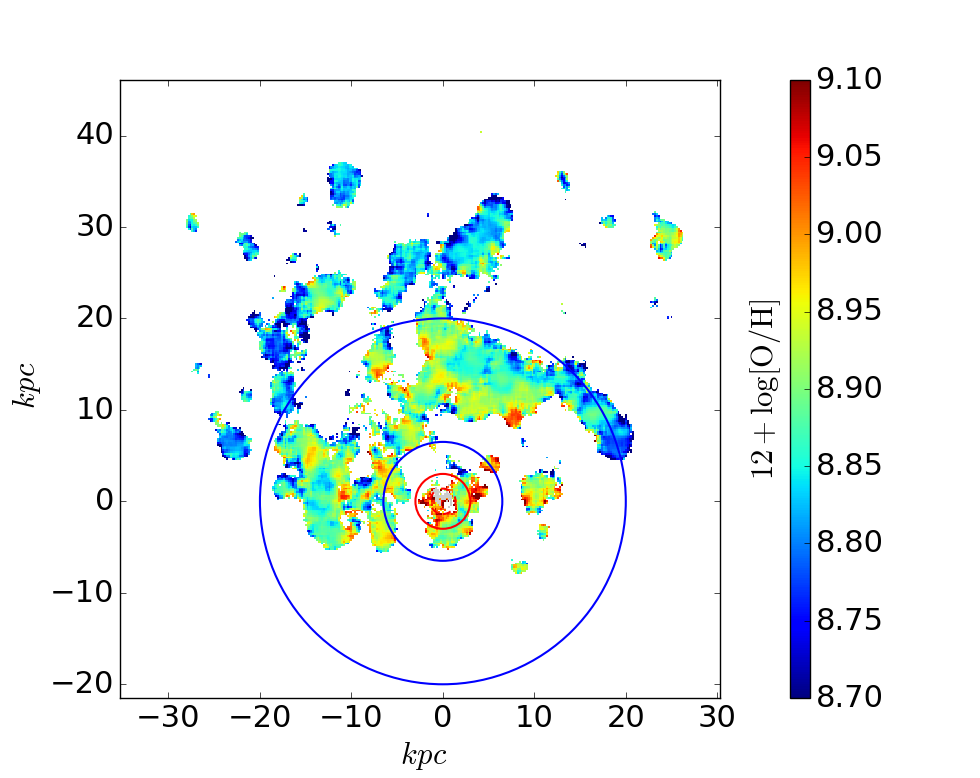}}
    \caption{Gas (top left, $\rm H\alpha$) and stellar (top right) kinematics in 
JO171: ionized gas is present only in the northern part of the 
galaxy, and forms elongated tails that are being stripped from the 
galaxy. The arrow points toward the cluster center (i.e. the Brightest 
Cluster Galaxy, BCG). Stars are instead distributed in a symmetric pattern 
around the galaxy center. The solid red circle delimits the spheroid and 
corresponds to the red solid line in Fig.\ref{fig:profile}, blue circles delimit the ring 
(also in blue in Fig.\ref{fig:profile}).  
The almost vertical line is the rotation axis of the stars, while the second inclined line ($\sim 37 \degr$) is the rotation axis of the central spheroid.
The 
lower left panel shows a zoom in the central region of the stellar 
kinematics map, where the stellar 
spheroid counter-rotation appears. The bottom right panel shows the 
gas metallicity.}
    \label{fig:kinematics}
\end{figure*}

By comparing the ionized gas and stellar kinematics (Fig.\ref{fig:kinematics}), we find that gas
and stars of the ring share the same rotation, on average, but have
quite different spatial distributions.
While the gas extends only towards the North, stars are evenly
distributed around the center. Gas and stellar velocities in the ring
are similar, and reach a maximum value of $\sim$100 $\rm km \, s^{-1}$. The rotation axes
of the two components are 
misaligned.  

The gas tails and the fact that the 
stellar kinematics is instead undisturbed are clear signs that the gas
is being stripped by a mechanism that affects the gas, but not the
stars, i.e. ram pressure due to the interaction with the
ICM. This is probably occurring approximately
along the direction traced by the arrow in the top left panel of
Fig.\ref{fig:kinematics} pointing towards the cluster BCG (Brightest Cluster Galaxy).

Interestingly, stars in the central spheroid counter-rotate with respect
to both stars and gas in the ring, with a rotation axis forming an
angle of $\sim$225 degrees, as can be seen in the zoom shown in the bottom
left panel of Fig.\ref{fig:kinematics}. Here the maximum projected velocity is
about 60 $\rm km \, s^{-1}$.  The counter-rotation of the spheroid with respect to the
ring is one of the major differences between JO171 and the Hoag's
galaxy.  We also note that the rotation axis of the spheroid is $\sim 37\degr$
 misaligned with respect to the one of the stellar ring.

The analysis of the kinematics clearly
points towards an accretion/merger scenario, since (i) stars in the
spheroid counter-rotate with respect to stars and gas in the ring,
(ii) there is no evidence of a bar, and (iii) the rotational axis of
stars in the ring and in the central spheroid are slightly misaligned.

The pure kinematical analysis cannot distinguish whether the
accretion was due to the slow disruption of a mostly gaseous companion,
or to cold accretion of gas along a filament,
and it is not able to date the event itself, unless
combined with the stellar population properties as done below.  

\subsection{Stellar populations}

\begin{figure*}
\centerline{\includegraphics[width=3in]{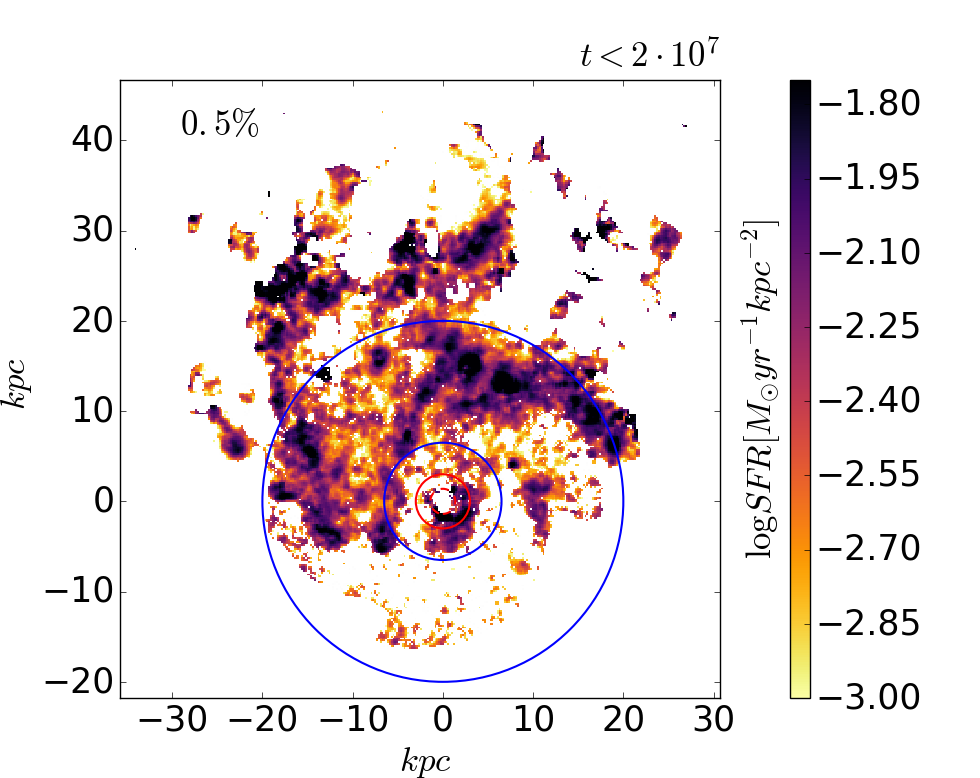}\includegraphics[width=3in]{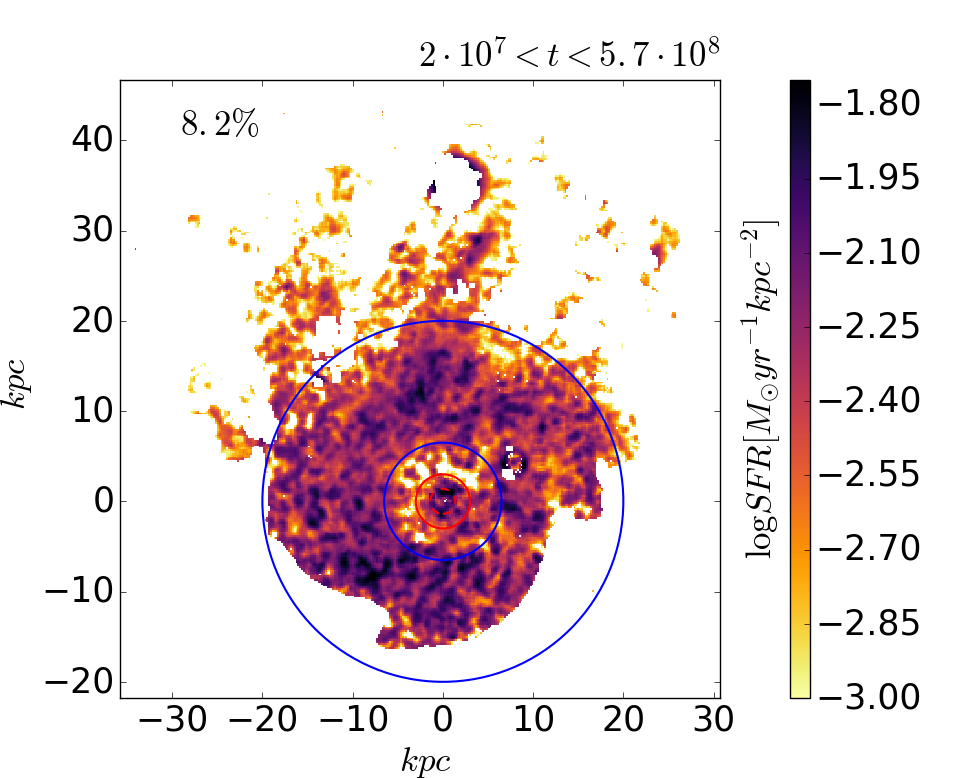}}
\centerline{\includegraphics[width=3in]{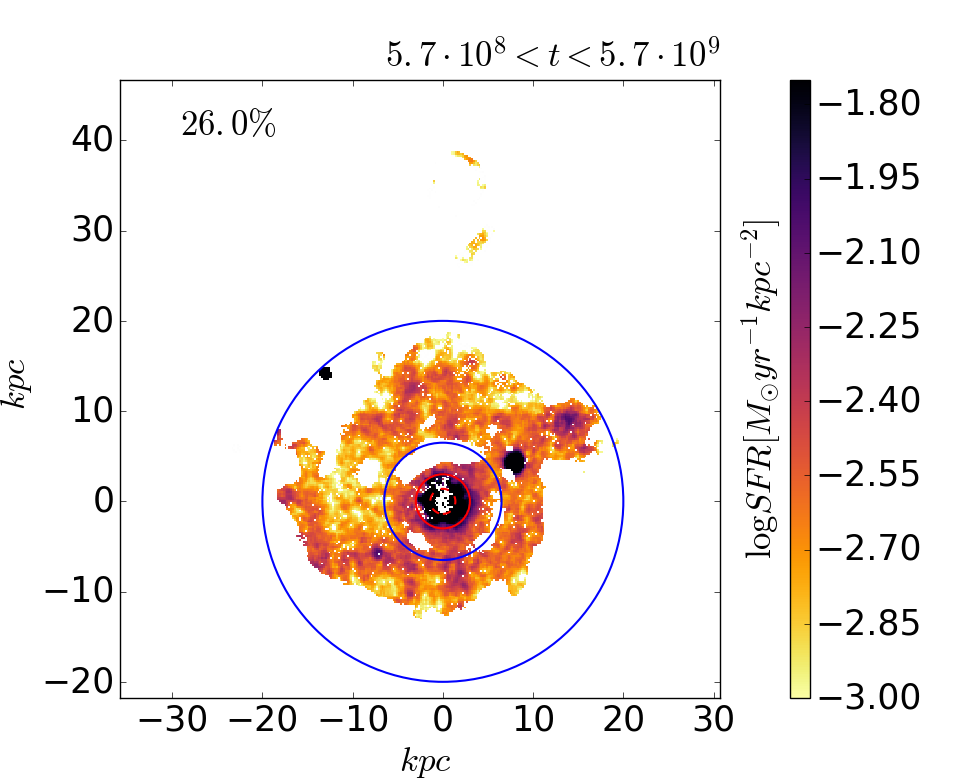}\includegraphics[width=3in]{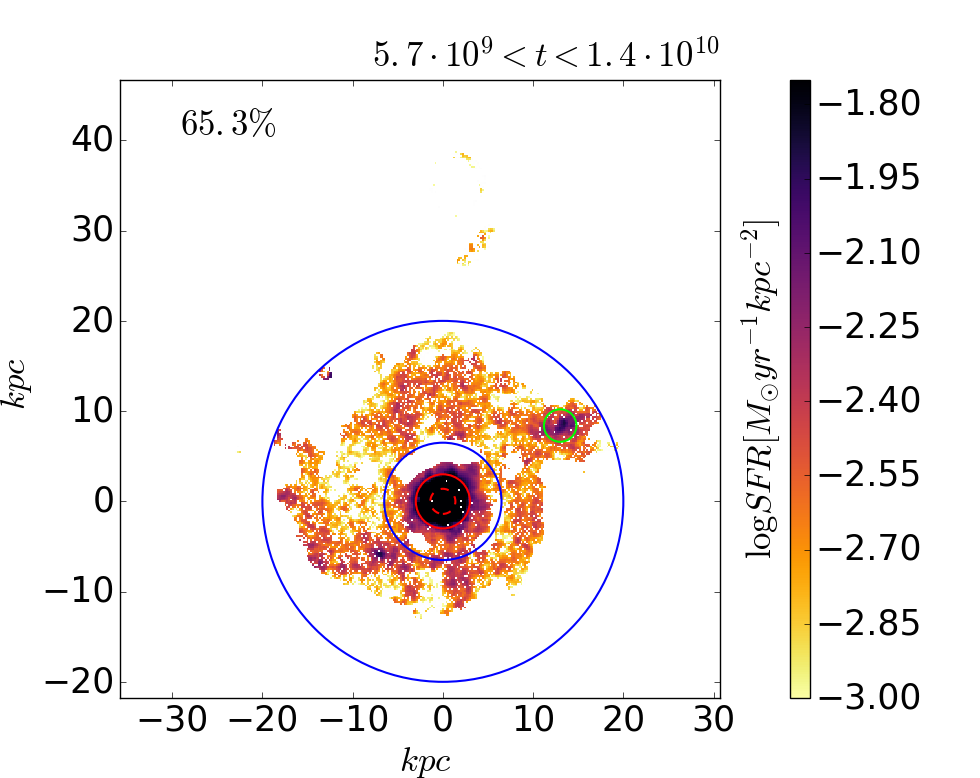}}
    \caption{Star formation rates surface densities in 4 age bins as 
derived from SINOPSIS: from top left to bottom right we show the 
ongoing SFR (ages younger than $2 \times 10^7$yr), the recent SFR (ages 
between $2 \times 10^7$ and $5.7 \times 10^8$yr), intermediate and old SFRs 
(with ages between $5.7 \times 10^8$ and $5.7 \times 10^9$ and older than $5.7 \times 10^9$yr,
respectively). The green circle in the bottom right panel shows the blob 
of most enhanced star formation within the ring in the oldest age 
bin.}
    \label{fig:sinopsis}
\end{figure*}

We further used the output from SINOPSIS to date the stellar
populations and reveal the star formation history experienced
by JO171.  Fig.\ref{fig:sinopsis} presents the logarithmic star formation rate density (in solar
masses per year per squared kpc) maps in four age bins characterizing
ongoing star formation (stellar age $<2 \times 10^7$ yr), recent star
formation (stellar ages between $2 \times 10^7$ and $<5.7 \times 10^8$yr), intermediate
age stars (between $5.7 \times 10^8$ and $5.7 \times 10^9$ yr) and old stars (older
than $5.7 \times 10^9$ yr).  

Globally, the fraction of present stellar mass (living stars and
remnants) formed in the four age bins is 0.5\%, 8.2\%, 26.0\% and
65.3\% (from younger to older age bins) of the total stellar mass,
i.e. about 91\% of the total mass of the galaxy was formed before the last 0.6 Gyr. 

Figure \ref{fig:sinopsis} shows that the star formation (SF) at early epochs was
concentrated in the spheroid, even though a low level of star
formation is seen also in the ring.  The SF in the ring increases with
time, peaking between $2 \times 10^7$ and $5.7 \times 10^8$ yr ago,
both in the Southern and Northern halfs of the ring. In contrast, the
ongoing SF (ages $< 2 \times 10^7$yr) is occurring only where the
ionized gas is still present, i.e. in the Northern half of the galaxy
and in the tails.

\begin{figure*}
\centerline{\includegraphics[width=3in]{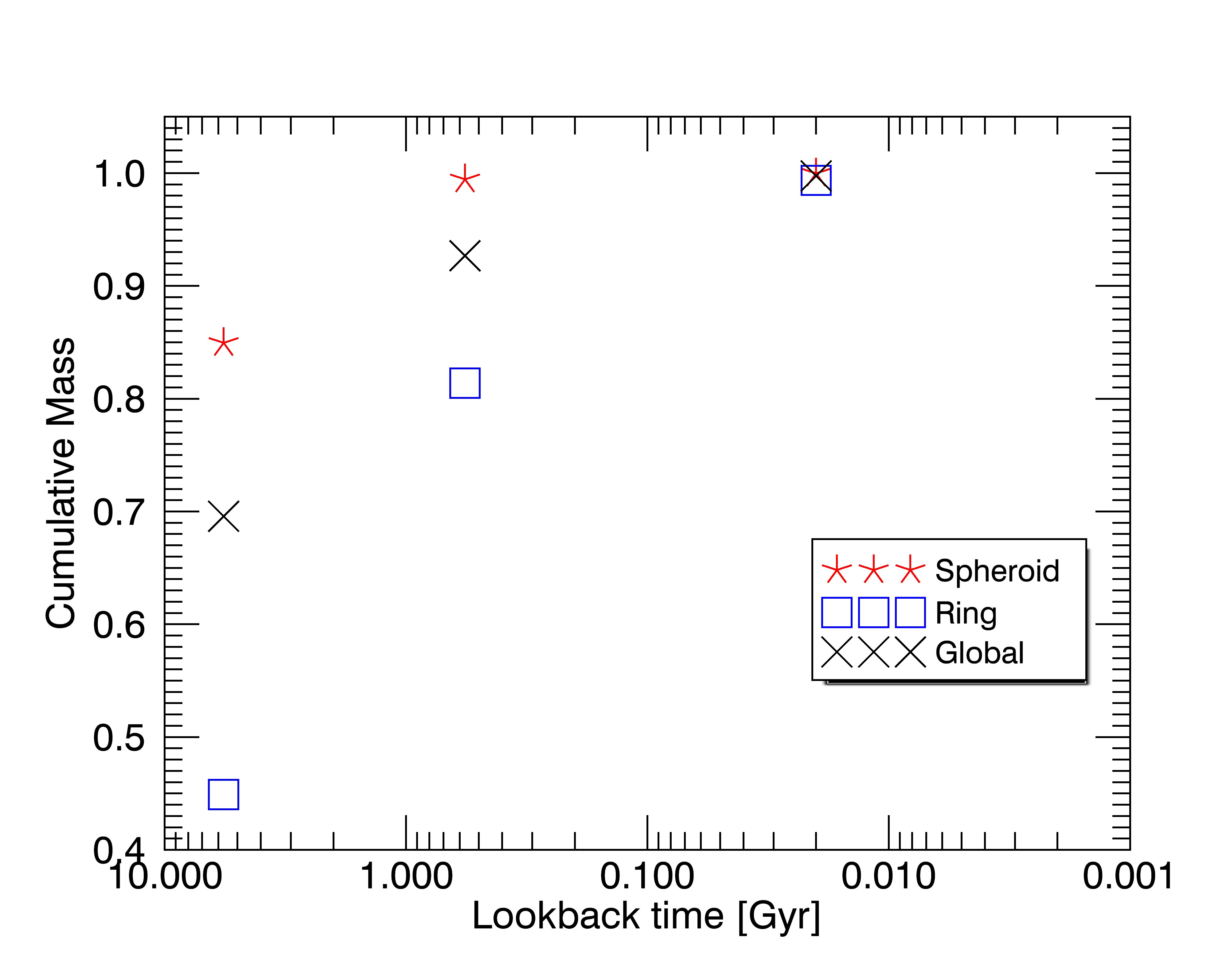}\includegraphics[width=3in]{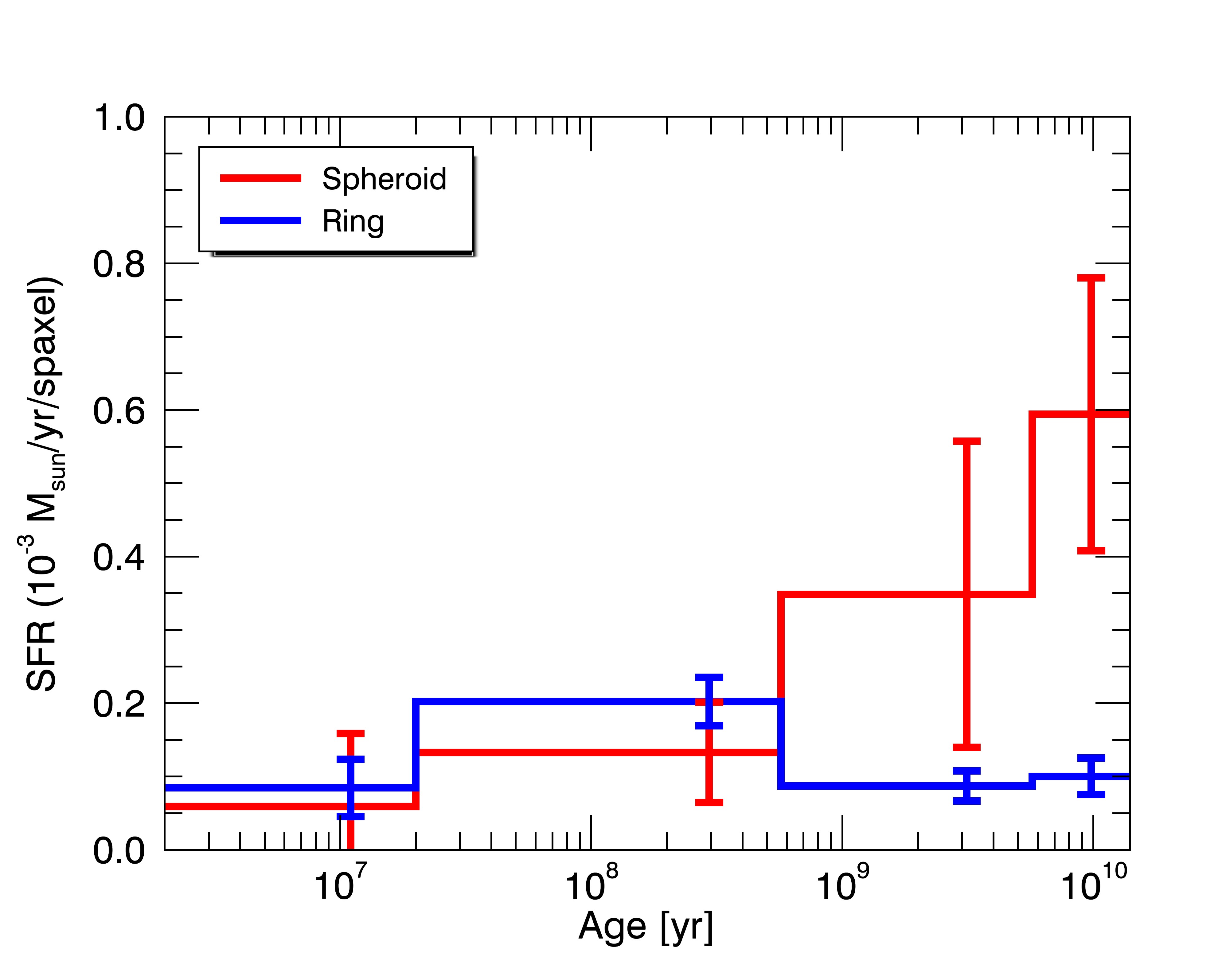}}
    \caption{Cumulative mass-age (left) and (right) average star 
      formation rate per spaxel-age 
      relations for JO171: the central spheroid is traced by the red 
      line/symbol and the ring by the blue. SFRs in the right panel 
      are in units of $10^{-3}$ solar masses per year per spaxel.}
    \label{fig:sfr}
\end{figure*}

In order to better understand the history of this galaxy, in Fig.\ref{fig:sfr}
we show the mass-age and star formation rate (SFR)-age relations.
The left panel shows how the stellar mass was assembled over cosmic time in the spheroid (red asterisks),
the ring (blue squares) and globally (black crosses). At the end of the oldest age bin,
i.e. 6 Gyr ago, almost 85\% of stars in the spheroid were already in
place, while only 45\% of the mass in the ring was born. About 0.6 Gyr
ago the spheroid was almost totally formed (99\%) while the ring had
$\sim$81\% of its current mass.  As better shown in the right panel of
Fig. \ref{fig:sfr} (with the same color coding), the star formation histories 
of the spheroid and the ring are quite different: while the spheroid is
characterized by a monotonously declining star formation, the ring
sees its star formation significantly enhanced in the recent age bin
(between 20 and 570 Myrs ago).

Overall, the analysis of the stellar population ages shows that 1) the
central spheroid formed at early epochs and had a monotonously
declining star formation rate; 2) the formation of the ring began 
during the oldest age bin, and the {\it average} SF of the ring was 
low and constant until 600 Myr ago. We notice that the mass locked
into old stars 
could be due either to a very low SFR
protracted for a very long time or to single episodes of high SF,
since the age resolution of spectrophotometric models is insufficient to
age-date individual SF episodes within each age bin.
3) the average SF in the ring strongly increased during the last 600 Myr,
forming 20\% of its current mass in this period. In the next section,
we will see that this increase is most
likely a result of a SF enhancement due to the impact of the galaxy
on the ICM when the galaxy entered the cluster.

\subsection{Environment}

JO171 is located in the cluster A3667, a very rich cluster with a
virial mass $M_{200}= 1.7 \times 10^{15} M_{\odot}$ \citep{Moretti+2017}, that
is dynamically very active:
it possesses a cold gas front in the center \citep{Vikhlinin+2001} and two bright diffuse
radio emission regions straddling the X-rays gas \citep{Rottgering1997}. A first analysis
of cluster substructures \citep{Owers+2009} also suggests that A3667 is the result of
two merging sub-clusters of similar masses in the plane of the sky,
probably occurred $\sim$1 Gyr ago.
JO171 lies at a projected clustercentric distance of $\sim$0.64 $R_{200}$
from the BCG, where $R_{200}$ is the 
radius at which the mean density of the cluster reaches 200 times the 
critical density and is expected to contain most of the virialized 
cluster mass.

\begin{figure*}
	\includegraphics[width=6in]{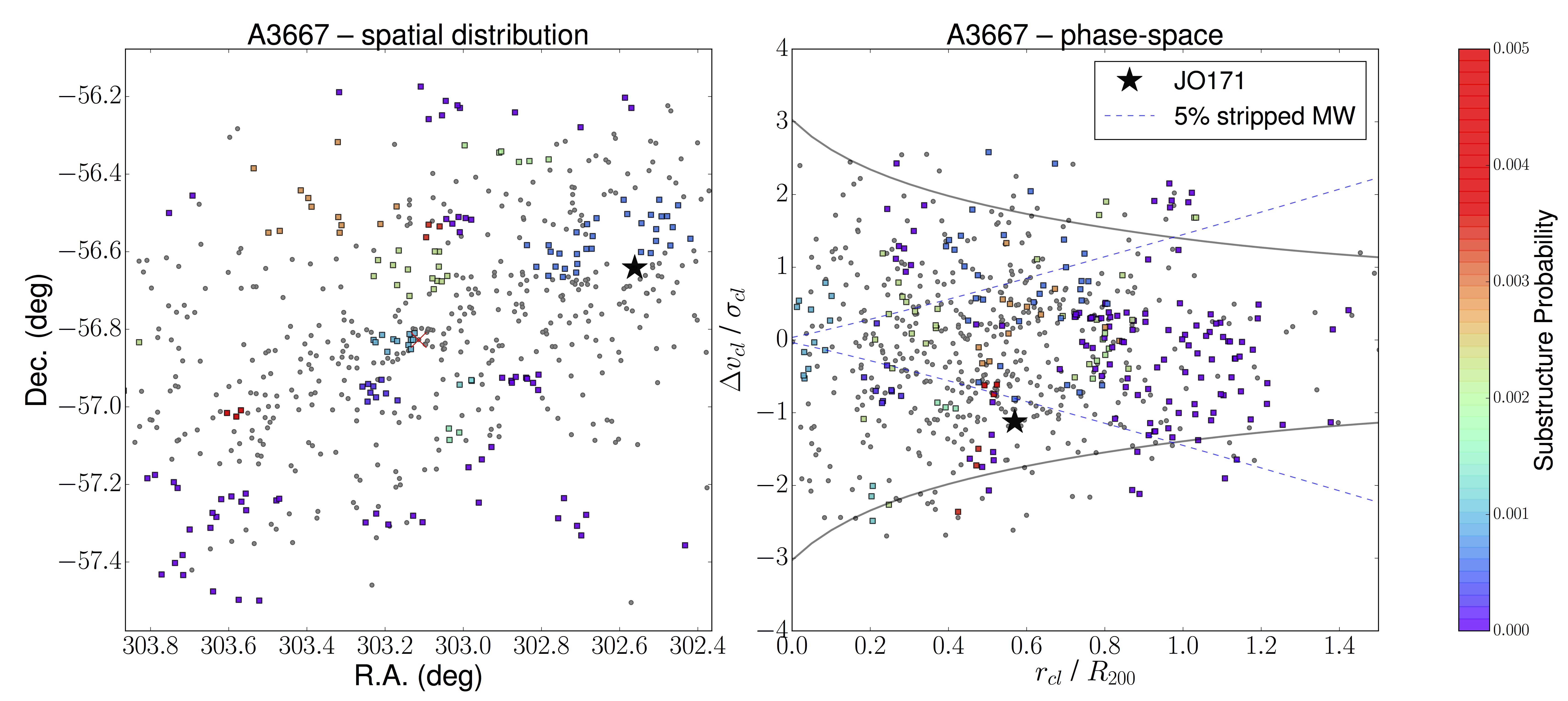}
    \caption{A3667 members and substructures: the left panel shows the 
spatial distribution of A3667 spectroscopic members (grey points) with 
superimposed the substructure members, color coded according to (1-p) 
where p is the probability to belong to a substructure. The right 
panel shows the same galaxies in the phase-space diagram. The black 
star shows the location of JO171. The dashed lines show the region 
where a Milky Way-like galaxy will lose 5\% of its gas due to ram 
pressure stripping. The continuous lines show the escape velocity in a NFW halo.}
    \label{fig:pps}
\end{figure*}

A deeper understanding of the complex galaxy-cluster dynamical
status could help linking the JO171's peculiar morphology to its
position in the cluster.  In particular, the dynamical analysis can
help clarifying if the galaxy belongs to a substructure infalling into
the cluster potential well or not.  The left panel of Fig.\ref{fig:pps} shows the
sky distribution of the spectroscopically confirmed cluster members, 
along with the identified substructures color-coded
according to the probability of being randomly extracted from a
Gaussian distribution with cluster typical values at the given
distance. Galaxies plotted with bluest colors have a
higher probability to belong to a substructure than the red
ones. 
After having assigned cluster membership using the KMM algorithm to
identify the main redshift peak, and then the Clean algorithm \citep{Mamon+2013} to refine it, the cluster substructures have been identified using all the spectroscopic
information available from the OmegaWINGS spectroscopy \citep{Moretti+2017}, complemented by literature data \citep{Owers+2009}, using
a dedicated algorithm described in \citet{Biviano+2017}.

While our algorithm does not assign JO171 (black star) to any
substructure, the galaxy is located close to some identified
substructures, so it is possible that there is also a physical -
albeit undetected - relation with them. More likely, this peculiar
galaxy was probably hit by a shock wave originated during the cluster
merger, as suggested by its location close to a "radio bridge"
identified NW of the cluster core \citep{Carretti+2012}.  

The right
panel of Fig.\ref{fig:pps} shows the projected clustercentric distance $r_{cl}$ plotted
against the line-of-sight velocity relative to the cluster velocity
$\Delta v_{cl}$. The resulting phase-space diagram (PPS) has
been normalized by the cluster size and velocity dispersion. For
reference, the escape velocity in a Navarro, Frenk \& White \citep{NFW1995} halo is
shown by the grey curves.  Cosmological simulations have shown that
galaxies of different time since accretion into the cluster occupy
different regions in projected phase-space, with the oldest cluster
members concentrating at low $r_{cl}$ and low $\Delta v_{cl}$, while recent infalls
having higher velocity dispersion and/or larger $r_{cl}$ \citep{Rhee+2017,Haines+2015,Oman+2013}.  The
location of JO171 in phase-space is
suggestive of recent accretion to the cluster, in agreement with the
hypothesis that the recent burst of star formation in the ring is
linked with the first impact with the cluster.

Since the work of \citealt{GG72} it is known that the ram pressure
exerted by the hot ICM on infalling galaxies is proportional to the
ICM density (that increases with decreasing $r_{cl}$ ), and the square of
the galaxy peculiar velocity. 
The stripping predicted from this simplified analytical prescription
coincides 
remarkably well with HI observations of cluster galaxies \citep{Jaffe2015,Yoon+2017}.  High-resolution cosmological
hydrodynamic simulations of cluster galaxies indeed predict that
galaxies can lose their cold gas during their first radial infall in
the cluster \citep{Cen+2014}. Simulations have also shown that
an enhanced ram pressure efficiency can be the
effect of ICM inhomogeneities, even in galaxies well
beyond the virial radius \citep{Tonnesen+2007}.

Assuming smooth ICM profiles, and Gunn \& Gott (1972)'s analytical
prescription of ram pressure stripping, it is possible to predict
the amount of gas stripping expected for infalling galaxies given
their location in PPS diagrams. In short, stripping occurs when the
intensity of ram pressure overcomes the anchoring force of the galaxy
(Jaff\'e et al. submitted).

The anchoring force of JO171 is difficult to model since this galaxy has two
components with different gas contents, and the gas in the ring 
is probably subject to a lower anchoring force than in a
normal disk.
Since JO171 peculiar morphology prevents a detailed modeling of the
anchoring force, a conservative estimate of the amount of gas
stripping can be obtained using as
proxy a Milky Way-like disk galaxy in a cluster with a velocity
dispersion $\sigma_{cl}$=1000 $\rm km \, s^{-1}$, similar to the A3667 velocity
dispersion (1010 $\rm km \, s^{-1}$), as in \citet{Jaffe2015}. This method has
already been used for GASP I \citep{gaspI}, GASP II \citep{gaspII}
, GASP III \citep{gaspIII} and GASP IV
\citep{gaspIV} papers.

At the location of JO171 in the
right hand plot of Fig.\ref{fig:pps}, we estimate that an infalling Milky
Way-like galaxy would have lost 5\% of its total gas mass via
ram-pressure (dashed blue lines, see simulations in 
Jaff\'e et al. submitted). This estimate is only a lower limit
for the following reasons: i) we expect that the ring of JO171, being
subject to a weaker anchoring force than the Milky Way, will suffer
from a more intense RP, that will strip the gas earlier during infall
(i.e. to the right of the dashed lines), ii) A3667 is a merging system
with an inhomogeneous ICM that can cause enhancement of RP, and iii)
the measured velocity and distance of the galaxy from the cluster
centre are projected values, and thus are lower limits to their real
values. Given the morphology of the tails, JO171 is probably moving
with respect to the cluster mostly within the plane of the sky, and
the real velocity relative to the ICM will be much higher than the
line-of-sight relative velocity observed.

In summary, the position of JO171 in phase-space indicates that the
galaxy must be undergoing ram pressure stripping by the ICM, and both
the presence of substructures in A3667 and the structural
configuration of this galaxy suggest that the pressure exerted on
JO171 is probably much higher than the one that can be computed 
with the standard assumptions.

\section{The origin of JO171}

Being a ring galaxy, JO171 offers the opportunity to study a secondary
event in the lifetime of a galaxy, and the detailed analysis of its
properties can cast light on the major/minor merger scenario as
opposed to gas accretion. In addition, JO171 experienced a tertiary
event, recently entering a rich cluster, and this gives us the unique
opportunity to study the effects of ram pressure on a fragile ring galaxy.

Regarding the formation of the ring, our analysis seems to rule out an internal origin (bar
induced resonances), given the absence of any bar
signature in our data. Moreover, the small, but detectable,
misalignment between the rotation axis of the stellar spheroid and the
outer disk seems to confirm an external
event in the galaxy's life.

We also discard the hypothesis that it is a collisional ring due to
the passage of a small galaxy through the disk, as this scenario is
unable to account for the formation of a counter-rotating component.

We are therefore left with only two scenarios: the gas rich merger and
the cold accretion along a filament of intergalactic medium.  Both
scenarios have been invoked to explain the most massive polar rings
\citep{Bekki1998,Iodice+2006,Spavone2013,Spavone2010}, and evidence of gas accretion from either filaments or
interactions with gas-rich satellites have been found in a few
galaxies \citep{Sancisi2008}.

In the major merger case, numerical simulations \citep{Jesseit+2007} are able to produce an old
and counter-rotating spheroid with a 1:1 retrograde merger only if
including a 10\% dissipative component. However, a certain amount of
gas inflow toward the galaxy center is expected \citep{Bournaud+2007}, and so is the
subsequent episode of star formation. Moreover, models predict that
the central galaxy will have a flattened luminosity profile, with
ellipticity e=1-b/a reaching 0.5-0.6 after 4 merger events. Neither
a significant young central stellar component, nor the flattening are
observed in JO171.

The analysis of the structural properties of JO171 spheroid confirms
that it is a fast rotator, therefore making the major merging
hypothesis very unlikely. 

Simulations of minor mergers \citep{Mapelli+2015} are able to produce the observed long-lived star
forming rings in many cases, but still induce a prolonged star
formation episode in the central galaxy. However, in these simulations
this effect is due to the presence of a bar in the main spheroid, that helps accumulating
material in the central region. Our data, therefore, can not rule out
a retrograde 1:2-1:5 merger as the origin of this galaxy.  If the
counter-rotating ring is the result of a relatively minor merger, we
might try to date the epoch of this merger based on the stellar
populations. 

Mergers with a gas-rich companion are known to trigger
bursts of star formation. From Fig.\ref{fig:sfr} it is apparent that $\sim$45\% of the
current stellar mass of the ring was already in place 6 Gyr ago. 

If we attribute the last burst of SF ($\leq$0.5 Gyr ago) to the entrance of
J0171 into the cluster and to the consequent RPS, this means that the
merger must have happened long before the entrance of JO171 in the
cluster.
Vice versa, if we attribute the last burst of SF to the merger, we would
date the first stages of the merger to $\sim$0.5 Gyr ago. This would imply that
the companion converted more than 81\% of the current stellar mass of
the ring into stars
before merging with the spheroid, and 
that JO171 entered the cluster and underwent a merger approximately at the
same time. 
However, this would also imply that
the companion should have been 
disrupted in a very short timescale (<0.5 Gyr), without leaving any
clear signatures of nucleus and tidal arms, which is
  unrealistic. 
Thus, if the ring of JO171 comes from a gas-rich merger, 
we expect that this merger happened more than 0.5 Gyr ago.

If this is what happened, we might be able to spot in the ring of
the system faint signatures of the merger remnant.  We identified in
particular one region in the ring where the star formation rate in the
oldest bin is enhanced (the green circle in the bottom right panel of
Fig. \ref{fig:sinopsis}), as a possible merger remnant, but its velocity turned out to
be $\sim 1000$ $\rm km \, s^{-1}$ larger than the galaxy one, thus excluding any link with JO171. 

Regarding the possibility that the gas is accreted from a nearby
companion,we verified that there are none within the MUSE field-of-view.

We also searched in our spectroscopic ($>90$\% complete at 
V=20) and photometric catalogs \citep{Moretti+2017,Gullieuszik+2015} for possible companions and gas donors. 
Following the study by \citet{Brocca+1997} we searched them within a 
region 5 times larger than the diameter of JO171's ring. 
We searched both galaxies with redshift brighter than V=20, and any
other candidate in the photometric catalog down to two magnitudes
fainter than JO171.
This search yields only two galaxies, 
 both 
fainter than JO171 (1.4 and 2.6 mag fainter in B than JO171).
They are both spectroscopically confirmed members lying at a projected 
distance of $\sim$90 kpc from JO171, and have line-of-sight relative velocity of about 
800 and 1350 $\rm km \, s^{-1}$ with respect to JO171. Given the high relative 
projected velocities we can exclude interactions among these galaxies,
also given the fact that the close environment of JO171 does not seem 
to be perturbed at the present epoch. Moreover, none of the three 
galaxies (JO171, and the two highlighted galaxies) seems part of any 
substructure (see \S3.4).

For all of these reasons, we consider the merger hypothesis
the least likely, and the cold gas
accretion along a filament hypothesis the most plausible scenario to explain
JO171 properties.  

Simulations \citep{Agertz+2009} predict that the accreted cold gas
settles into a large disk-like system, which then fragments and forms
stars in clumpy regions.
In our case, this accretion should have started long
before the galaxy accreted onto the cluster, as the old age of the
ring stars demonstrates. Such rings can survive as long as 2-8 Gyrs
\citep{Maccio2006},
and have masses similar to the central spheroid.  

In such a scenario, the metallicity of the gas in the ring 
is expected to be
lower than
the standard one for the galaxy total galaxy mass or luminosity.
This result has already been 
confirmed in polar ring galaxies \citep{Spavone2010}. 

After having removed the stellar
component from the MUSE spectra, we derived the gas metallicity
shown in Fig.\ref{fig:kinematics}, bottom right panel.  The gas has a
median metallicity of 12+log(O/H) $\sim$ 8.87,
that is slightly lower than the expected 9.06 for a normal
mass-metallicity relation \citep{Tremonti+2004}
and lower than the average 9.1-9.2 value for its B-band luminosity
\citep{Spavone2010}. This effect, however, is small for JO171, at
the 2 $\sigma$ level. This is
because the ring and the spheroid have similar masses, while the
effect should be much more visible if the ring were only a lower
fraction of the total mass/luminosity.
Moreover, the gas shows no hints for structured metallicity gradients,
that would be expected in case of normal metal enrichment due to
stellar populations aging in a disk galaxy.

Therefore, we propose a scenario of gas accretion at early epochs
(first $\sim 6$ Gyr), followed by a more recent star formation enhancement
(as demonstrated by the higher SFR in the second youngest age bin) due to
the interaction with the environment, at the beginning of the ram
pressure effect. The ram pressure is also responsible for the complete
stripping of the gas from the South region, and for the tentacles
towards the North, where we observe the ongoing star formation.

\section{Summary}

Within our ongoing ESO MUSE GASP Program, we have studied the
characteristics and the evolutionary history of JO171, a peculiar
$\sim 3 \times 10^{10} M_{\odot}$ ring
galaxy in the dense environment of the cluster A3667.  This galaxy
resembles the well known Hoag's galaxy \citep{Finkelman+2011} and is
constituted by a central spheroid surrounded by a ring of young
stars/gas. Its ring gas is currently being stripped due to the ram-pressure
exerted by the ICM on the infalling galaxy.  Our data show that
this galaxy underwent two transformations that dramatically changed
the course of its evolution: the formation of the ring, most probably early on in
its history, and the more recent stripping of its ring gas.
It is the first time that a peculiar ring galaxy similar to the Hoag's object is observed in a
dense cluster in the act of being transformed 
by environmental effects.

The availability of MUSE integral field spectroscopy allowed us to
investigate its peculiar kinematics and to unravel the properties of
its stellar populations with unprecedented details.  As the Hoag's
galaxy, JO171 does not show evidence for a central bar, strongly
suggesting that the formation of the ring is due to an external cause.
Moreover, at odds with the Hoag's galaxy, the central spheroid of JO171 
counter-rotates with respect to the ring, confirming that a secondary event
must be at the origin of the ring.

JO171 currently displays long tails of ionized gas towards the north,
while its southern half is totally devoid of gas. Both of these
characteristics are due to ram pressure stripping acting approximately
in the direction opposite to the cluster center.

The joint analysis of the stellar/gas
kinematics and the stellar populations properties shows that,
while the spheroid is generally old and characterized by a
monotonously declining star formation history, the formation of the
ring stars began early on, with an average SFR low and constant
until $\sim$ 600 Myr ago, then there was an increase in star formation activity
during the last 500 Myr due to ram pressure upon cluster infall. 

Both gas rich minor mergers and accretion of gas
along a filament can in principle be at the origin of the ring, 
but the lack of any evidence for a merger remnant in the MUSE data
favors the second hypothesis.
Moreover, the analysis of the MUSE field of view and of the surrounding
environment seems to rule out the presence of companions and possible
gas donors around JO171.

JO171 gives us a unique opportunity to peer into the 
mechanisms that drive the formation of peculiar ring 
galaxies and their evolution in dense environments.  Our study
supports a scenario in which peculiar ring galaxies 
can be formed through a gas accretion event along a filament, as it was already 
suggested for the isolated Hoag's galaxy \citep{Finkelman+2011} and polar ring 
galaxies \citep{Spavone2013,Jore1996}.  
If this is the case either the ring is able to survive 
for several Gyr before being destroyed, or it is slowly formed, as the 
stellar population ages suggest.  
In the case of JO171, the fate of the ring is driven by the fact that,
when the galaxy interacts with the cluster ICM, the gas gets stripped and
the star formation in the ring is switched off, leaving an aging and fading
stellar-only ring.

\section*{Acknowledgements}

We thank the anonymous Referee.
This work is based on observations collected at the European Organisation for Astronomical Research in the Southern Hemisphere under ESO program 196.B-0578. This work is also based on observations taken with the AAOmega spectrograph on the AAT and the OmegaCAM camera on the VLT. This work made use of the KUBEVIZ software, which is publicly available at http://www.mpe.mpg.de/$\sim$dwilman/kubeviz/. We acknowledge financial support from PRIN-INAF 2014 and from the INAF PRIN-SKA 2017 program 1.05.01.88.04. M. M. acknowledges financial support from INAF through grant PRIN-2014-14 and from the MERAC Foundation.
J.F. acknowledges financial support from a UNAM-DGAPA-PAPIIT IA104015 grant, Mexico. 
B.V. acknowledges the support from an Australian Research Council Discovery Early Career Researcher Award (PD0028506).




\bibliographystyle{mnras}
\bibliography{gasp5} 

\bsp	
\label{lastpage}
\end{document}